\documentclass[conference]{IEEEtran}
\IEEEoverridecommandlockouts
\usepackage{cite}
\usepackage{amsmath,amssymb,amsfonts}
\usepackage{algorithm}
\usepackage{algpseudocode}
\usepackage{graphicx}
\usepackage{tabularx}
\usepackage{textcomp}
\usepackage{xcolor}
\usepackage{gensymb}
\usepackage{subcaption}
\usepackage[hidelinks]{hyperref}
\usepackage[capitalise]{cleveref}
\creflabelformat{equation}{#2#1#3}
\def\BibTeX{{\rm B\kern-.05em{\sc i\kern-.025em b}\kern-.08em
    T\kern-.1667em\lower.7ex\hbox{E}\kern-.125emX}}

\makeatletter
\newcommand{\linebreakand}{%
  \end{@IEEEauthorhalign}
  \hfill\mbox{}\par
  \mbox{}\hfill\begin{@IEEEauthorhalign}
}
\makeatother

\begin{document}

\title{A Study on Optimization Techniques for Variational Quantum Circuits in Reinforcement Learning}

\author{
\IEEEauthorblockN{1\textsuperscript{st} Michael Kölle} 
\IEEEauthorblockA{\textit{LMU Munich} \\
Munich, Germany \\
michael.koelle@ifi.lmu.de}
\and
\IEEEauthorblockN{2\textsuperscript{nd} Timo Witter} 
\IEEEauthorblockA{\textit{LMU Munich} \\
Munich, Germany \\
t.witter@campus.lmu.de}
\and
\IEEEauthorblockN{3\textsuperscript{rd} Tobias Rohe} 
\IEEEauthorblockA{\textit{LMU Munich} \\
Munich, Germany \\
tobias.rohe@ifi.lmu.de}
\linebreakand
\IEEEauthorblockN{4\textsuperscript{th} Gerhard Stenzel} 
\IEEEauthorblockA{\textit{LMU Munich} \\
Munich, Germany \\
gerhard.stenzel@ifi.lmu.de}
\and
\IEEEauthorblockN{5\textsuperscript{th} Philipp Altmann} 
\IEEEauthorblockA{\textit{LMU Munich} \\
Munich, Germany \\
philipp.altmann@ifi.lmu.de}
\and
\IEEEauthorblockN{6\textsuperscript{th} Thomas Gabor} 
\IEEEauthorblockA{\textit{LMU Munich} \\
Munich, Germany \\
thomas.gabor@ifi.lmu.de}
}

\maketitle

\begin{abstract}
Quantum Computing aims to streamline machine learning, making it more effective with fewer trainable parameters. This reduction of parameters can speed up the learning process and reduce the use of computational resources. However, in the current phase of quantum computing development, known as the noisy intermediate-scale quantum era (NISQ), learning is difficult due to a limited number of qubits and widespread quantum noise. To overcome these challenges, researchers are focusing on variational quantum circuits (VQCs). VQCs are hybrid algorithms that merge a quantum circuit, which can be adjusted through parameters, with traditional classical optimization techniques. These circuits require only few qubits for effective learning. Recent studies have presented new ways of applying VQCs to reinforcement learning, showing promising results that warrant further exploration. This study investigates the effects of various techniques --- data re-uploading, input scaling, output scaling --- and introduces exponential learning rate decay in the quantum proximal policy optimization algorithm's actor-VQC. We assess these methods in the popular Frozen Lake and Cart Pole environments. Our focus is on their ability to reduce the number of parameters in the VQC without losing effectiveness. Our findings indicate that data re-uploading and an exponential learning rate decay significantly enhance hyperparameter stability and overall performance. While input scaling does not improve parameter efficiency, output scaling effectively manages greediness, leading to increased learning speed and robustness.
\end{abstract}


\section{Introduction}
Quantum computing is currently transforming the research landscape by offering the potential for accelerating computation exponentially, making it a promising solution for today's most complex computational challenges. Algorithms like Grover's allow faster searches in unsorted databases, while Shor's algorithm can factorize large numbers exponentially faster \cite{nielsen2001quantum,meyer2022survey}. Particularly, the field of machine learning, especially reinforcement learning (RL), benefits from quantum computing due to its potential to reduce trainable parameters and accelerate learning processes. RL aims to enable an agent to find optimal strategies through interaction with an environment, leading to notable successes in games like Chess and Go, surpassing human champions, and in teaching robots to learn directly from camera feeds \cite{campbell2002deep,silver2016mastering,levine2016end}. However, many of today's RL achievements face challenges of exponentially growing computational and memory requirements with problem complexity. Neural networks (NNs) can be substituted by quantum circuits, promising fewer trainable parameters, improved learning speed and sample efficiency \cite{meyer2022survey}. However, in today's NISQ era, quantum circuits face their own challanges: low qubit counts, noise and a lack of quantum error correction \cite{meyer2022survey}. Therefore, we focus on variational quantum circuits, which show resilience to noise and can achieve learning successes with few qubits \cite{chen2020variational}. While VQCs have sometimes underperformed compared to NN algorithms, enhancements like input and output scaling and data re-uploading have matched or exceeded NN performance in certain cases \cite{QRL_jerbi2021parametrized,skolik2021layerwise,QRL_skolik2022quantum,perez2020data_re-uploading}. However, the specific contributions of these improvements and their reproducibility remain less explored. We use an proximal policy optimization (PPO) algorithm, replacing the actor NN with a modifiable VQC, aiming to evaluate methods through this quantum proximal policy optimization (QPPO) algorithm with reproducible results and a fair comparison to classical PPO with an equivalent parameter count. We additionally introduce a new technique that uses exponential learning rate decay. For our experiments, we use Pennylane's \cite{bergholm2018pennylane} cost-effective quantum simulation and two environments from the OpenAI gym library \cite{brockman2016openai}: a modified Frozen Lake environment \cite{chen2020variational} to evaluate data re-uploading and output scaling and the Cart Pole environment to evaluate alternative parameter initializations \cite{gaussinit_zhang2022escaping}, input scaling and exponential learning rate decay.

The rest of the paper is structured as follows: In \cref{sec:Related-Work} we review current research and related works which employ VQCs for RL problems. \cref{sec:Methods} details the QPPO algorithm used, \cref{sec:experimental_setup} describes the experimental setup. We present and discuss our results in \cref{sec:Results} and \cref{sec:Discussion}. Lastly, discuss these findings and conclude our work in \cref{sec:Conclusion}.

\section{Related Work}\label{sec:Related-Work}
In this section, we provide a brief summary of recent studies on quantum reinforcement learning (QRL) that utilize VQCs. We focus on key advancements and techniques in the field, including Quantum Q-Learning, Quantum Policy Gradient Methods, Quantum Soft Actor-Critic, Quantum Advantage Actor-Critic, and Quantum Proximal Policy Optimization, shedding light on their significance and applications.

\emph{Quantum Q-Learning:} Chen \textit{et al.} \cite{chen2020variational} pioneered the use of VQCs in RL to approximate the action-value function, showcasing the effectiveness of hybrid architectures in the NISQ era for quantum machine learning. Building on this, Lockwood and Si \cite{QRL_VA_lockwood2020reinforcement} extended the approach to continuous state spaces and further demonstrated that VQCs could match the performance of NNs in RL settings. In a subsequent study, they tackled an Atari video game environment using a hybrid NN-VQC setup, highlighting challenges in attributing learning progress between the components \cite{QRL_VA_lockwood2021playing}. Skolik \textit{et al.} \cite{QRL_skolik2022quantum} improved upon Chen \textit{et al.}'s work, utilizing data re-uploading and output scaling in Frozen Lake and Cart Pole environments. They emphasized the importance of output interval scaling for value-based VQCs, noting challenges with increasing circuit complexity. Chen \textit{et al.} \cite{QRL_VA_chen2022variational} explored gradient-free optimization of a VQC for Q-learning through evolutionary strategies, circumventing the barren plateau problem and indicating potential for significant advancements in QRL.

\emph{Quantum Policy Gradient Methods:} Diverging from Q-learning, Jerbi \textit{et al.} \cite{QRL_jerbi2021parametrized} applied a VQC to learn the policy of a quantum policy gradient algorithm, incorporating data re-uploading and input scaling to enhance performance and achieve a quantum advantage in a specialized environment. Sequeira \textit{et al.} \cite{QRL_sequeira2023policy} also used a VQC for a parametrized policy, achieving results comparable to more parameter-heavy NNs, though the specific factors contributing to this success remain unclear.

\emph{Quantum Soft Actor-Critic:} Lan \cite{QRL_lan2021variational} developed one of the first quantum versions of the soft actor-critic algorithm for continuous action spaces, testing two approaches with varying degrees of hybrid NN-VQC usage. A similar approach with data re-uploading was explored by Acuto \textit{et al.} \cite{QRL_acuto2022variational} for controlling a virtual robotic arm, showing improvements in performance with hybrid models.

\emph{Quantum Advantage Actor-Critic:} Kölle \textit{et al.} \cite{Hgog2023} implemented an advantage actor-critic algorithm with a VQC architecture, suggesting modifications to circuit design but highlighting the necessity of hybrid VQC approaches for complex environments.

\emph{Quantum Proximal Policy Optimization:} Kwak \textit{et al.} \cite{QRL_PPO_kwak2021introduction} integrated a VQC into a proximal policy optimization (PPO) algorithm, replacing only the Actor network. This work demonstrated that VQCs can exceed random behavior, offering a foundational step for future QRL research. Hsiao \textit{et al.} \cite{QRL_hsiao2022unentangled} proposed a unique hybrid system, using 1-Qubit rotations without entanglement, indicating potential for quantum-inspired RL algorithms even on classical hardware.

\section{Method}\label{sec:Methods}
\begin{figure}
    \centering
    \includegraphics[width=\linewidth]{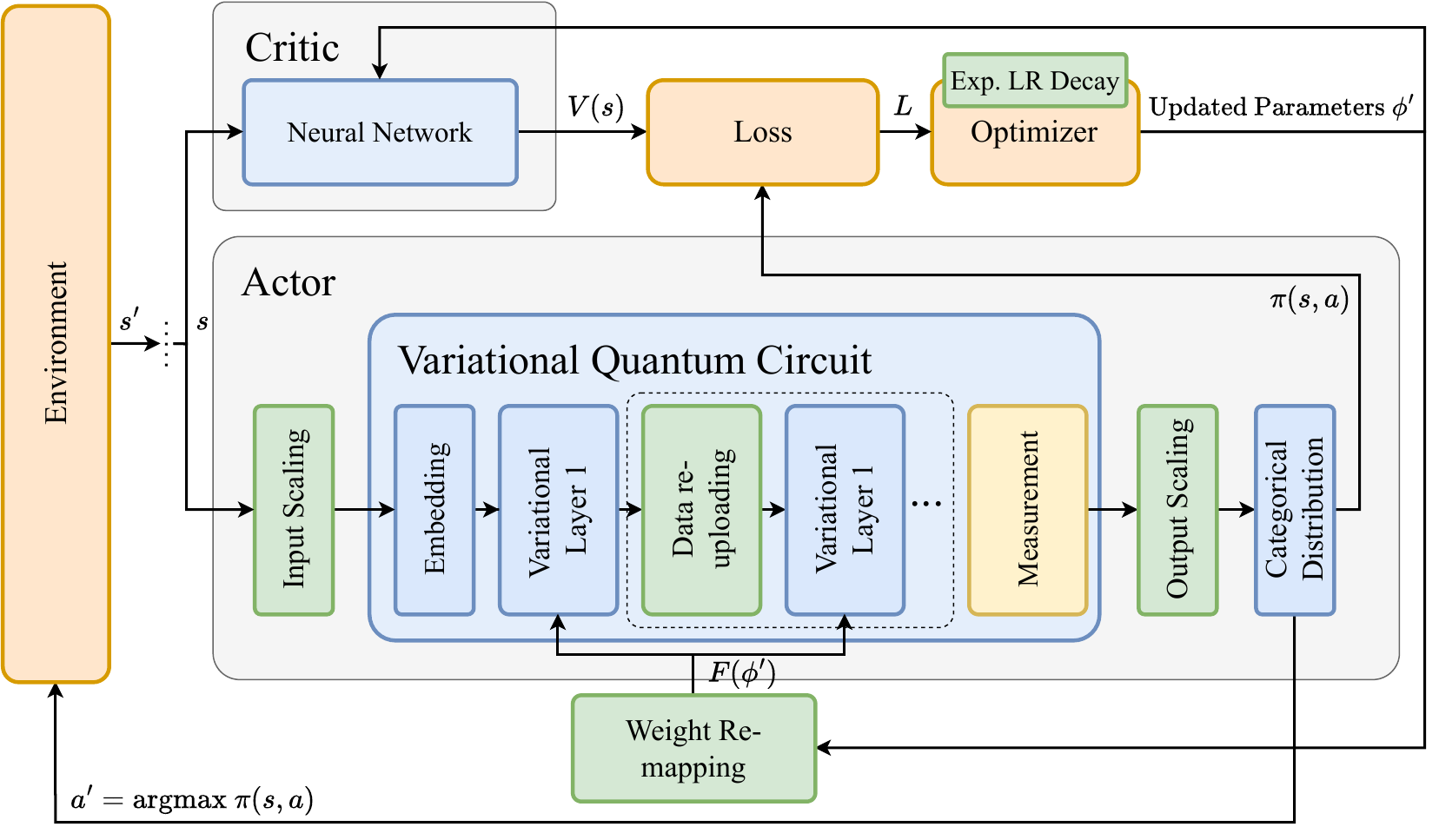}
    \caption{Overview of our methodology. Green boxes denote the techniques we investigated in this work.}
    \label{fig:overview}
\end{figure}

The objective of this work is not to introduce a new algorithm surpassing classical approaches, but to evaluate methods of improvement for their application in a QPPO algorithm. This involves testing whether these methods can indeed enhance the performance of the VQC relative to the number of parameters used and whether this suffices to outperform a NN with a comparable number of parameters. To ensure reliable results, we replace only the actor of the PPO algorithm with a VQC (or a small NN for comparison), while maintaining a standard NN with two hidden layers of 64 nodes each (totaling 5313 parameters for Frozen Lake and 4545 for Cart Pole) for the critic. Given the critic's significantly larger parameter set, designed for more complex environments, it likely learns nearly optimally with standard hyperparameters and thus requires no further consideration. A overview of our methodology is depicted in \cref{fig:overview}.


We describe the logic of the (Q)PPO in \cref{subsec:PPO_logic} and the VQC architecture for the actor in \cref{subsec:circuit}.We then introduce exponential learning rate decay in \cref{subsec:exp_lr_sced}, list initialization methods in \cref{subsec:initialisation}, and explain input and output scaling in \cref{subsec:outscale} and \cref{subsec:input_Scaling}, respectively.

\subsection{Quantum Proximal Policy Optimization Algorithm}\label{subsec:PPO_logic}
Both classical and QPPO share the same algorithm logic (\cref{alg:PPO}), based on an already well-optimized version \cite{PPO_implement}. The vectorized environment is initialized, and the initial state for each environment is set via the reset() function. Actor and critic, along with their parameters and respective Adam optimizers, are initialized. Batches for states $s$, actions $a$, log probabilities $logp$, rewards $r$, dones $d$, and values $v$ are prepared with null values before starting the training loop for updates. During each update, 128 steps are executed in each environment using the current policy of the actor, storing the corresponding values in the batch. Afterward, advantages and returns are calculated from the values, rewards, and dones. Finally, four update cycles are conducted, where pointers to the batch are randomly shuffled and divided into four minibatches for loss calculation. The combined loss function is minimized to update the parameters \cite{PPO_implement,PPO_schulman2017proximal}. The architecture of classical and quantum PPO remains analogous, except for the actor selection: a NN for the former and a VQC for the latter, as described in \cref{subsec:circuit}.

\begin{algorithm}

\begin{algorithmic}
\State Initialize environment and set next state $s_{next}$
\State Initialize optimizer, parameters for the critic, the actor, and the input scaling and/or output scaling if used
\State Initialize batch for states $s$, actions $a$, logprobs $logp$, rewards $r$, dones $d$ und values $v$
\For{each update} \do:
\If{using exponential $lr$ schedule}
    \State Calculate and set current actor $lr$
    \If{(Local) input scaling}
        \State Set input scaling $lr$ to current actor $lr$
    \EndIf
\EndIf
\For{number of environment steps}\do:
    \State $s_{step}$ == $s_{next}$
    \For{i in range(number of envs)}\do:
        \State Get $a_{step_{i}}$, $logp_{step_{i}}$ from actor
        \State Get $v_{step_{i}}$ from critic with $s_{step_{i}}$
    \EndFor
    \State $s_{next}$, $r_{step}$, $d_{step}$ = $envs.step(a_{step})$
\EndFor
\State Get next value $v_{next}$
\State Calculate \textit{advantages} und \textit{returns} for $r$, $d$, $v$ and $v_{next}$
\For{each update epoch} \do:
    \State shuffle batch indizes, make minibatches
    \For{each minibach} \do:
        \State Initialise batch for new $logp^{new}$, $v^{new}$, $S$
        \For{each mimibach step} \do:
            \State Get $logp^{new}_{step}$, $S_{step}$ from the actor, 
            \State Get $v^{new}_{step}$ from the critic with $s_{step}$
        \EndFor
        \State Calc. \textit{ratio} with $\pi_{old}=e^{logp}$ and $\pi=e^{logp^{new}}$
        \State Normalise \textit{advatages} at minibatch level
        \State Calc. policy loss using \textit{advatages} and \textit{ratio}
        \State Calc. value loss with batches of $v$, $v^{new}$, \textit{returns}
        \State Calc. entropy loss as $S$.mean()
        \State Calc. the combined loss
        \State Optimize parameters
    \EndFor
\EndFor
\EndFor
\end{algorithmic}
\caption{(Quantum) Proximal Policy Optimization}\label{alg:PPO}
\end{algorithm}

\subsection{Variational Quantum Circuit Architecture}\label{subsec:circuit}
\begin{figure*}[hpbt]
    \centering
    \subfloat[Data re-uploading circuit for Frozen Lake]{
    \includegraphics[width=0.54\linewidth]{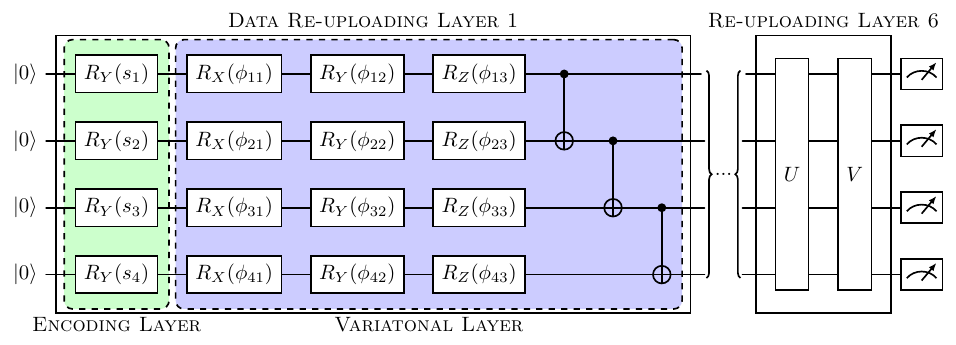}
    \label{fig:circuit-FL}
    }
    \hfill 
    \subfloat[Data re-uploading circuit for Cart Pole]{
    \includegraphics[width=0.415\linewidth]{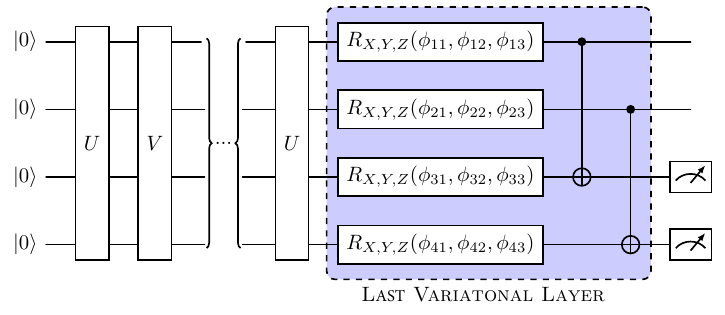}
    \label{fig:circuit-CP}
    }
    \caption{Data re-uploading circuits. \textbf{(a)} For Frozen Lake: Utilizes 6 standard data re-uploading layers, each with an encoding and a variational layer featuring \emph{C-NOT} entanglement. The encoding layer inputs the binary state $s$ digits, and the variational layer processes rescaled parameters $\phi = \pi \times \tanh(\theta)$, with \emph{Pauli-Z} measurements on all qubits. \textbf{(b)} For Cart Pole: Also employs 6 data re-uploading layers, similar to the Frozen Lake circuit but differs in the last layer. The variational layer uses rescaled parameters $\phi = \pi \times \tanh(\theta)$. The encoding layer, without input scaling, inputs state dimensions $z_{RES}=rescale(s)$ or, with input scaling, the values $z_{IS}=\pi \times \tanh(\lambda \times s)$, utilizing input scaling parameters $\lambda$ and the environmental state $s$, with \emph{Pauli-Z} measurements on the last two qubits.}
\end{figure*}

We selected a hardware-efficient VQC approach with $4$ qubits, proven effective in prior studies \cite{kandala2017hardware_efficient,leone2022practical_hardware_efficient}. Given the unresolved question of how architecture influences learning success, the literature features diverse approaches for quantum actor-critic algorithms. Our circuit builds on the foundation laid by \cite{QRL_PPO_kwak2021introduction}, demonstrating the learning capability of a QPPO with this VQC as the actor. It employs Y-rotation ($R_Y(\phi)$) for state encoding, \emph{X}, \emph{Y}, and \emph{Z} rotations in each variational layer, and cascading (non-circular) \emph{C-NOT} quantum gates for entanglement \cite{QRL_PPO_kwak2021introduction} (\cref{fig:circuit-FL}). For the Cart Pole environment, this VQC entangles only qubit $1 \rightarrow 3$ and  $2 \rightarrow 4$ in the final step, performing measurements only on the last two qubits, not all like \cite{QRL_PPO_kwak2021introduction} (\cref{fig:circuit-CP}).

Since rotations in the Hilbert space are $\pi$-periodic, meaning $R_{X/Y/Z}(\phi + 2\pi) = R_{X/Y/Z}(\phi)$ for any angle $\phi$, we use a \emph{tanh weight-remapping} technique \cite{kolle2022improving} to limit the input for each rotation in the variational layer to the interval $]-\pi, \pi[$.

\subsubsection{Encoding}\label{subsec:encoding}
To encode the 16 states of the Frozen Lake environment (\cref{subsec:Frozen Lake}) into the VQC's four qubits, we use binary encoding, which converts the discrete state number into a binary number matching the qubits' count (e.g., $3 \rightarrow 0011$ or $15 \rightarrow 1111$). We then multiply each binary value by $\pi$ and use it as the angle for an encoding rotation on one of the qubits \cite{chen2020variational,QRL_skolik2022quantum}.

For Cart Pole, we apply angle encoding commonly used for continuous state spaces. Here, we directly use the value of a state space dimension as the input angle for a parametrized rotation, after applying a rescaling function \cite{QRL_skolik2022quantum,QRL_lan2021variational}. This rescaling function maps each dimension of the continuous state space to a $2\pi$ interval (typically $[-\pi, \pi]$), preventing the encoding rotations' $2\pi$-periodicity from translating different states into the same rotational angle. Typically, dimensions with fixed intervals are linearly scaled to $[-\pi, \pi]$ by dividing by their upper limit, while potentially infinite dimensions are constrained using a $\pi \times \tanh()$ function. For the Cart Pole state tuple $s = (x_{C}, v_{C}, \phi_{P}, v_{P})$, the rescaling $z$ is defined as:

\begin{equation}
z_{RES} = \pi \times ({\frac{x_{C}}{4.8}},\ \tanh(v_{C}),\ {\frac{\phi_{P}}{0.418}},\ \tanh(v_{P}))
\end{equation}

This scaling accounts for the cart position $x_{Cart}$ within $[-4.8, 4.8]$, cart velocity $v_{Cart}$ in $[-\infty, \infty]$, pole angle $\phi_{Pole}$ in $[-0.418, 0.418]$ radians, and pole velocity $v_{Pole}$ in $[-\infty, \infty]$. Many studies use this or similar scalings, e.g., a $2 \times \arctan()$ function instead of $\pi \times \tanh()$ (Kölle \textit{et al.} \cite{Hgog2023}); as per \cite{kolle2022improving}, the $\tanh$ approach appears superior to $\arctan$ for initial learning efficiency.

\subsubsection{Data Re-uploading}
Additionally, we employ data re-uploading \cite{perez2020data_re-uploading}, a method that increases the incorporation and entanglement of environmental state information into the circuit, thus improving the VQC's stability and learning capability. Instead of a single encoding layer at the circuit's outset, we placed one before each variational layer. With a chosen depth of six, this results in six alternating encoding and variational (or re-uploading) layers \cite{perez2020data_re-uploading,QRL_jerbi2021parametrized,QRL_skolik2022quantum,QRL_lan2021variational}.

\subsubsection{Measurements and Action Selection}\label{subsec:measurement_and_action_selection}
In Frozen Lake, we perform a \emph{Pauli-Z} measurement on each qubit, while in Cart Pole, measurements occur only on the last two qubits. The measurement outcomes then correspond to the probabilities of each action. However, as the VQC's output ranges between $[-1,1]$, a normalization function is required before these values can serve as probabilities for drawing random actions from a categorical distribution. Without output scaling, we apply the standard normalization (for the shifted interval), resulting in probabilities $p$:
\begin{equation}
p_i = \frac{C_i^{out} + 1}{\sum_{j=1}^n (C_j^{out} + 1)}
\end{equation}
where $n$ is the number of actions, $i = 1,...,n$, and $C^{out}$ is the VQC output. With output scaling, we utilize a parametrized softmax:
\begin{equation}
p_i = \frac{e^{\beta C_i^{out}}}{\sum_{j=1}^n e^{\beta C_j^{out}}}
\end{equation}
with the trainable output scaling factor $\beta$ \cite{QRL_jerbi2021parametrized}.

\subsection{Alternative Circuit Architectures}\label{subsec:alt_arch}
\begin{figure*}[hpbt]
    \centering
    \subfloat[Kölle \textit{et al.} \cite{Hgog2023} re-uploading circuit for Frozen Lake]{
    \includegraphics[width=0.47\linewidth]{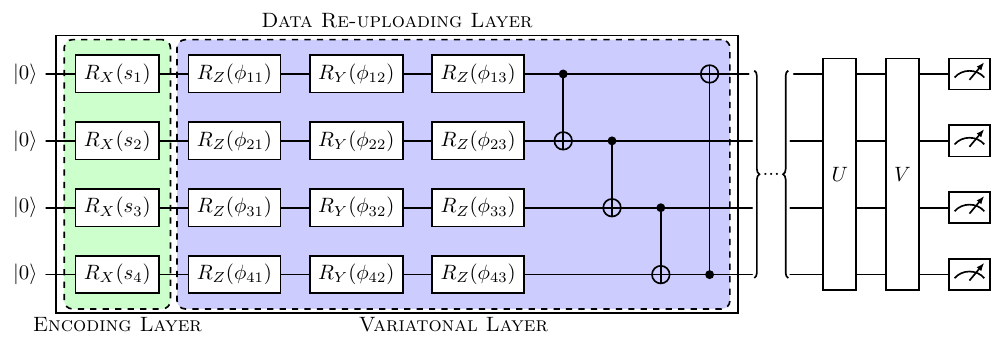}
    \label{fig:Hgog-circuit}
    }
    \hfill
    \subfloat[Jerbi \textit{et al.} \cite{QRL_jerbi2021parametrized} re-uploading circuit for Frozen Lake]{
    \includegraphics[width=0.47\linewidth]{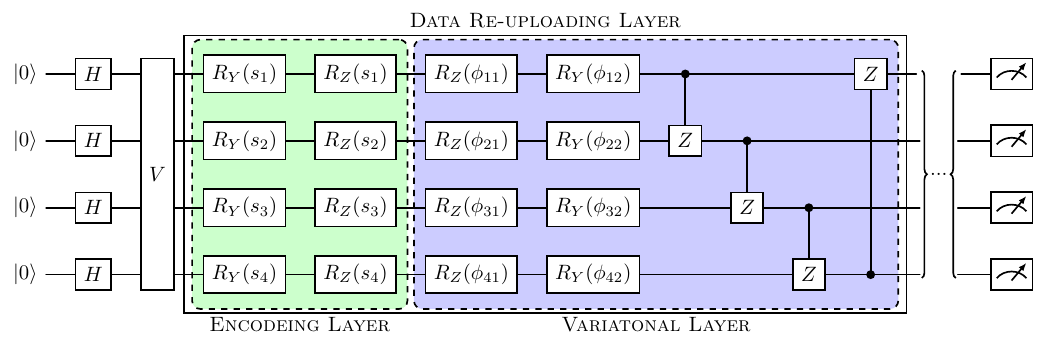}
    \label{fig:Jerbi-circuit}
    }
    \caption{Data re-uploading circuits for Frozen Lake. (a) Following Kölle \textit{et al.} \cite{Hgog2023}, this circuit uses 6 data re-uploading layers with parametrized rotations and CNOT entanglement. (b) Based on Jerbi \textit{et al.} \cite{QRL_jerbi2021parametrized}, it employs 8 data re-uploading layers with parametrized rotations and ontrolled-Z entanglement, with a Hadamard gate and an additional variational layer preceding them.}
\end{figure*}

Given the unresolved effect of VQC architecture on learning success, we introduce two additional circuit designs from the literature for comparison with our approach. As mentioned in \cref{subsec:circuit}, the standard circuit (based on Kwak \textit{et al.} \cite{QRL_PPO_kwak2021introduction}) features \emph{X}, \emph{Y}, and \emph{Z} rotations in each variational layer and data re-uploading. Kölle \textit{et al.} \cite{Hgog2023} explored a similar approach their work on quantum A2C, replacing $R_Y(\phi)$ encoding with $R_X(\phi)$ and employing \emph{Z}, \emph{Y}, and \emph{Z} rotations with circular entanglement in the variational layer. Their findings suggest that the \emph{Z}, \emph{Y}, \emph{Z} rotation sequence outperforms the \emph{X}, \emph{Y}, \emph{Z} sequence, which warrants further investigation. This alternative also incorporates data re-uploading for a fair comparison (\cref{fig:Hgog-circuit}).

Another noteworthy architecture, due to its uniqueness, comes from Jerbi \textit{et al.} \cite{QRL_jerbi2021parametrized}. It prepares each qubit with a Hadamard gate, with variational layers comprising \emph{Z} and \emph{Y} rotations and circular \emph{CZ} (\emph{Controlled-Z}) gates for entanglement. The encoding layers consist of \emph{Y} and \emph{Z} rotations. Uniquely, this approach starts with a variational layer followed by re-uploading layers, each including an encoding and a variational layer (\cref{fig:Jerbi-circuit}). Testing these alternative approaches in Frozen Lake aims to clarify the architectural influence relative to performance enhancement methods.

\subsection{Exponential Learning Rate Decay}\label{subsec:exp_lr_sced}
A common challenge with VQCs in RL environments is the need for precisely tuned hyperparameters to achieve noticeable learning progress. Preliminary testing consistently showed that the QPPO's VQC requires a very low learning rate ($lr$) to stabilize towards the end of training. This scenario presents a dilemma: too low initial lrs lead to slow early learning, while higher $lr$s, although beneficial initially, often prevent convergence within a competitive timeframe compared to classical PPO.

To enhance hyperparameter stability and minimize the steps required for convergence early in hyperparameter testing (thereby saving on computational and time resources), we introduce an exponentially decaying learning rate for the VQC. The concept of an adaptive learning rate, known as learning rate annealing in classical machine learning \cite{annealing_nnakamura2021learning}, typically involves a linear decrease over the learning process to zero. While sufficient for neural networks and often outperformed by a sigmoid function for deep NNs \cite{annealing_nnakamura2021learning}, these approaches are less suitable for a VQC, which may benefit from a learning rate ten to a hundred times larger in early phases than in mid to late phases. Our preliminary experiments suggest that the learning rate must decrease relatively quickly to minimize the slow early phase and ensure a stable learning process, making exponential decay the most straightforward choice for this study. We also set a fixed half-life (HL) for reproducibility and a lower bound for the learning rate, excluding it from exponential decay to guarantee late-phase learning capability. The exponentially decaying learning rate is defined as:
\begin{equation}
lr_t = \frac{lr_{Start} - lr_{End}}{2^{\frac{t}{HL}}} + lr_{End}
\end{equation}
where $lr_t$ is the learning rate at the current time step $t$ that decays with half-life (HL), $lr_{Start}$ is the initial learning rate, and $lr_{End}$ is the final or minimum learning rate.

$lr_{Start}$ can be set to the value that achieves the best initial results, even if likely too high for later phases. $lr_{End}$ is chosen to be low enough for stable convergence but still sufficient for significant learning progress within the available time. The starting learning rate $lr_{Start}$ will exponentially approach the ending rate $lr_{End}$ with a half-life of HL, gradually reducing the update steps as learning progresses. 

\subsection{Initialization Strategies}\label{subsec:initialisation}
Since suboptimal initializations can lead to very small gradients and even vanishing gradients, we investigate the impact of different intervals on performance. Additionally, we experiment with Gaussian initialization to prevent barren plateaus at the optimization's start \cite{gaussinit_zhang2022escaping} (the initializations are summarized in \cref{tab:Inits}). For the parametrized rotations of the VQC (with trainable parameters $\theta$), we employ a \textit{tanh} rescaling \cite{kolle2022improving} of the form $R_{X/Y/Z}(\pi \times \tanh(\theta))$, requiring the parameters $\theta$ to be initialized as random values between $-1$ and $1$ (or another chosen interval) and then applying the \textit{arctanh} to all. For Gaussian initialization, we use a Gaussian distribution with a standard deviation of $1$ and a mean of $0$. No \textit{arctanh} is needed for this initialization, as the drawn values already fall within the correct magnitude (since the 99\% limit of the normal distribution is at 3, and $\tanh(3) = 0.995$).

\begin{table}[htbp]
\begin{tabularx}{\linewidth}{|l|l|X|}
\hline
Name         & Method                           & Interval                             \\ \hline
\hline
Standard     & Uniform (arctanh) & {[}-1;1{]}                                 \\
Small        & Uniform (arctanh) & {[}-0.11; -0.01{]} $\cup$ {[}0.01; 0.11{]} \\
Medium       & Uniform (arctanh) & {[}-0.75; -0.25{]} $\cup$ {[}0.25; 0.75{]} \\
Large        & Uniform (arctanh) & {[}-0.99; -0.59{]} $\cup$ {[}0.59; 0.99{]} \\
Gaussian     & Gaussian\protect\footnotemark & {[}-0.995; 0.995{]} (99\% boundary)                   \\
Clipped Gaussian & Gaussian Clipped Magnitude & {[}-0.99; -0.01{]} $\cup$ {[}0.01; 0.99{]} \\ 
\hline
\end{tabularx}
\caption{Initializations for VQC parameters. $^1$Gaussian distributions use a standard deviation of 1 and a mean of 0}
\label{tab:Inits}
\end{table}

\subsection{Output Scaling}\label{subsec:outscale}
Unlike neural networks, a quantum circuit's output cannot assume arbitrary values and is limited to the $[-1,1]$ interval, potentially causing issues. A contemporary solution to this problem is output scaling, which applies a trainable parameter to each output dimension of the circuit. While implementations of scaling are similar, it's essential to distinguish between two approaches based on their function: a trainable output interval \cite{QRL_skolik2022quantum} for value-based methods and trainable greediness scaling \cite{QRL_jerbi2021parametrized} for policy-based methods. This work employs the latter version for the actor's output. Incorporating the critic as a VQC and using the trainable output interval exceeds this study's scope, but results from \cite{skolik2021layerwise} are referenced.

\subsubsection{Trainable Output Interval}\label{subsubsec:trainable_output_range}
For value-based approaches (or the actor in an actor-critic scheme), precisely approximating a value or action-value function with the circuit's limited output interval is challenging. Although only the relative sizes of the values matter for action selection in value-based algorithms, the value function update typically uses squared deviation to align the circuit's output with observed values as closely as possible. If the interval's maximum is significantly smaller than the real values, many would be capped at the highest possible value, rendering the algorithm incapable of distinguishing between good and bad states (or state-action pairs). Skolik \textit{et al.} \cite{QRL_skolik2022quantum} also highlights the issue of choosing an overly large interval, as it decreases the distinguishability of smaller values, particularly impeding early learning progress. They demonstrate that a dynamically adjustable interval offers a clear advantage over a static one, presenting an efficient resolution to this dilemma.

\subsubsection{Greediness Scaling}\label{subsubsec:greedyness_scale}
In PPO, the quantum circuit's normalized measurements form the basis for a categorical distribution. However, unlike in a standard PPO where a neural network's output serves directly as the log probability for the distribution, a VQC's output is confined to between $-1$ and $1$. For instance, in a two-dimensional output scenario (e.g., in Cart Pole), the maximal probability difference could be represented by the tuple (-1,1), leading to unnormalized probabilities of $(e^{-1},e^{1}) = (0.368, 2.718)$ or normalized probabilities of $(12\%, 88\%)$. This implies that even if the circuit identifies the optimal action, there's at least a 12\% chance of choosing the inferior action. Since many environments can end a run with a single incorrect step, and flawless action selection is crucial for success, this approach is clearly insufficient.

Jerbi \textit{et al.} \cite{QRL_jerbi2021parametrized} introduces a trainable output scaling solution, multiplying a simple factor by the VQC's output before normalization with softmax, thus leading to greedier action selection for values greater than 1 (e.g., $2 \times (-1,1) \approx (e^{-2},e^{2}) \approx (1.8\%;98.2\%)$). This factor can start at 1 and be progressively increased by an optimizer, controlling the speed of the transition from exploration to exploitation phase depending on the scaling learning rate. Since Jerbi \textit{et al.} \cite{QRL_jerbi2021parametrized} did not specify whether a shared parameter (global scaling) for all actions or an individual parameter for each action (local scaling) should be used, both scenarios are tested. One might expect local scaling, with four parameters in the Frozen Lake environment, to have an advantage since only the actions \emph{right} and \emph{down} are necessary to reach the goal, and a simple bias towards these actions could contribute to learning success. However, as both actions are equally required in Cart Pole, only global scaling is applied. The initial implementation without scaling, by normalizing the VQC's output without softmax (thus $(-1,1) \approx (0\%,100\%)$), simplifies the approach but linearly approximates a discrete action selection, which could be inadequate for achieving nuanced learning progress (as discussed in \cref{subsec:measurement_and_action_selection}).

\subsection{Input Scaling}\label{subsec:input_Scaling}
Input scaling, recently introduced by Jerbi \textit{et al.} \cite{QRL_jerbi2021parametrized}, enhances a VQC's expressiveness without adding additional layers. It employs a trainable parameter $\lambda$ for each encoding rotation, multiplied by the state value before its insertion into the encoding rotation (or rescaling for the rotation). As mentioned in \cref{subsec:encoding}, angle encoding requires rescaling to limit each dimension of the continuous state space to a $2\pi$ interval (typically $[-\pi, \pi]$) to prevent the $2\pi$-periodicity of encoding rotations from translating different states into the same rotational angle. With input scaling allowing dimensions to assume potentially infinite values, a \textit{tanh} function is used for all dimensions, resulting in a $R_Y(\pi \times \tanh(\lambda \times \theta))$ rotation for all encoding layers of the approach. To understand the impact of this (compared to manual rescaling) on the VQC's learning capability, global input scaling is also tested, employing a single parameter for each environmental dimension (each qubit) instead of one per encoding rotation. Although Jerbi \textit{et al.} \cite{QRL_jerbi2021parametrized} showed significant improvement with this parameter-intensive input rescaling, it's critical to assess whether these additional parameters were cost-effectively utilized. Since the primary goal is to reduce the number of parameters rather than layer depth, it remains to be determined if this approach can outperform a VQC with a similar parameter count but fewer layers. The initialization of input scaling in Cart Pole was chosen similar to the manual rescaling from \cref{subsec:encoding}, dependent on the dimensions being encoded: $\lambda_{x_{Cart}}=\frac{1}{4.8},\ \lambda_{v_{Cart}}=1,\ \lambda_{\phi_{Pole}}=\frac{1}{0.418},\ \lambda_{v_{Pole}}=1$; the scaling learning rate is set to the VQC learning rate, or for the global version, the output scaling learning rate.

\section{Experimental Setup}\label{sec:experimental_setup}
In this section, we go into detail about the environments (\cref{sec:Environments}), baselines for comparing results (\cref{sec:baseline}), training procedures (\cref{sec:training}), and metrics (\cref{sec:M}) used in our experiments. Conducted experiments are described in \cref{sec:experiments}.

\subsection{Test Environments}\label{sec:Environments}
We evaluate our algorithm in two gymnasium environments: a modified Frozen Lake environment \cite{chen2020variational} to evaluate data re-uploading and output scaling and the Cart Pole environment to evaluate alternative parameter initializations \cite{gaussinit_zhang2022escaping}, input scaling and exponential learning rate decay.

\subsubsection{Deterministic Frozen Lake}\label{subsec:Frozen Lake}


For our experiments we use the deterministic Frozen Lake environment with modified rewards, one of the simplest environments feasible for evaluating architectural decisions despite the resource-intensive simulation process. 
The environment features a partially frozen lake with the task to find the shortest path from start (\emph{S}) to goal (\emph{G}) without falling into holes (\emph{H}), terminating the run. The discrete state space is represented by a number for each grid ($0$ to $15$ for a $4x4$ world), with four possible movement actions: \emph{left}, \emph{down}, \emph{right}, or \emph{up}. To encourage finding the fastest path, we use the modified reward version, where reaching the goal yields $+1$, falling into a hole $-0.2$, and each step taken $-0.01$. The maximum reward is therefore $0.95$, with a local minimum of $-0.22$ for the fastest fall into a hole, relevant for interpreting our preliminary results in the appendix.

\subsubsection{Cart Pole}\label{Cartpole}


Our experiments in the Cart Pole environment (version 1) are intended to evaluate input encoding in a continuous state space and verify the QPPO algorithm's effectiveness in more complex environments. In the Cart Pole environment a pole is attached by an joint to a cart that moves along a track. The goal is to prevent the pole from falling over by balancing it through moving the cart left or right. The four-dimensional continuous state space consists of the cart's position ($x_{Cart}$) within $[-4.8, 4.8]$, its velocity $v_{Cart}$, the pole's angle $\phi_{Pole}$ within $[-24\degree, 24\degree]$, and the tip's velocity $v_{Pole}$. Only two actions, \emph{left} and \emph{right}, are available. The episode ends if the cart's position exceeds $(-2.4, 2.4)$ or the pole's angle surpasses $12\degree$, with a reward of $+1$ for each timestep survived, up to a maximum duration of $500$.


\subsection{Baselines}\label{sec:baseline}
We compare the quantum PPO against two baselines: a random agent and a classical PPO. The classical PPO shares the same algorithm logic as its quantum counterpart (\cref{alg:PPO}) except that it uses a neural network as function-approximator.

Both the PPO and QPPO's critic use a neural network with two hidden layers of 64 nodes each, resulting in 5313 parameters for Frozen Lake and 4545 for Cart Pole due to different input dimensions. Actor networks for Frozen Lake utilize one hidden layer with 3 or 4 neurons, totaling 67 and 88 trainable parameters (Plots reference these parameter counts). 
For Cart Pole, the actor networks have two hidden layers, which have $5x5$, $6x5$, $6x6$, leading to 67, 77 and 86 parameters, respectively. 
We choose orthogonal initialization for the weights and constant initialization with zero for biases as suggested in \cite{PPO_implement}.


\subsection{Training}\label{sec:training}
We implemented the classical PPO's actor and critic using the deep learning library \emph{PyTorch}, and the VQC using the quantum framework \emph{Pennylane}\cite{bergholm2018pennylane}, which also simulated the quantum computers for the learning phase. The training was conducted using a cluster of Linux machines, featuring an Intel(R) Core(TM) i5-4570 CPU. On average, executing 150,000 timesteps with the QPPO took about 20 hours.

Although the actor and critic share a common loss for optimization, we use a separate Adam optimizer (with $\epsilon = 10^{-5}$) for each parameter group. All experiments were run with at least three different seeds (some more sensitive ones with five). Each update consists of four cycles with four mini-batches (\cref{tab:general_hyperparams}). We employed four parallel environments in a synchronous vector environment, executing 128 timesteps in each iteration ($batchsize = 512$). The actor learning rate, output scaling learning rate, and used initialization were optimized in both environments over 150,000 timesteps in Frozen Lake and over 500,000 in Cart Pole (\cref{tab:merged_hyperparams}). Due to time constraints, we allocated approximately the same amount of time for all hyperparameter optimization runs. Preliminary experiments and hyperparameters can be found in the appendix.

\subsubsection{Critic Architecture}
All PPO versions utilize a classical critic with a learning rate of $2.5 \times 10^{-4}$ and two hidden layers of 64 nodes each. This setup is considered nearly optimal for straightforward problems like Frozen Lake and Cart Pole, particularly due to its high parameter count relative to the actor.

\subsubsection{Classical Actor}
The classical PPO's actor for Frozen Lake uses a reduced neural network with one hidden layer of 3 to 5 nodes, and for Cart Pole, two hidden layers of 5 to 7 nodes, aligning the number of trainable parameters with the VQC actor. Of course, larger networks will perform even better in these benchmarks.

\subsubsection{Quantum Actor}
For QPPO, we employ a VQC as described in \cref{subsec:circuit} with 6 layers (depth). Without output scaling, the VQC's output values are normalized and used as probabilities for a categorical distribution. With output scaling, these are multiplied by the scaling factor before normalization. Data re-uploading and uniform initialization for parameters in Frozen Lake, and a smaller initialization for Cart Pole, are used according to \cref{subsec:initialisation}.

\subsection{Metrics}\label{sec:M}
We measure experiments by the average results across runs relative to the total timesteps and trainable parameters. The mean cumulative reward during training over an update cycle (512 timesteps) in four parallel environments is defined as the result, smoothed with an exponentially weighted moving average (with $\alpha = 0.3$ for Frozen Lake and $\alpha = 0.015$ for Cart Pole for results; $\alpha = 0.05$ for hyperparameter tests). Robustness and stability of convergence are also considered in interpreting results and selecting hyperparameters.

\subsection{Experiments}\label{sec:experiments}
We conducted experiments to assess the methods presented in \cref{sec:Methods}, utilizing the QPPO across four experiments in two different environments. Initially focusing on Frozen Lake, we tested four versions of the VQC, incorporating techniques like data re-uploading, global output scaling, and an exponentially decaying learning rate, to evaluate the impact of each method on overall performance.
Subsequently, we investigated local output scaling's potential to enhance performance beyond greediness-scaling by biasing actions less commonly needed in the environment (specifically \emph{left} and \emph{up}.
We also compared the influence of VQC architecture, pitting a fully modified approach against two others enhanced with the three methods from literature.
In the final series, we examined input scaling in the Cart Pole environment. To ensure a fair comparison despite introducing many new parameters, we tested configurations with 4 and 5 layers (including re-uploading), comparing them to the conventional approach and a version with global input scaling.
To illustrate performance, we benchmarked all four approaches against the classic PPO, adjusting the actor network to match the parameter count of the VQCs.

\section{Results}\label{sec:Results}
\begin{figure*}[hpbt]
  \centering
  \subfloat[QPPO vs. classical PPO in Frozen Lake over three runs.]{
   \includegraphics[width=0.30\linewidth]{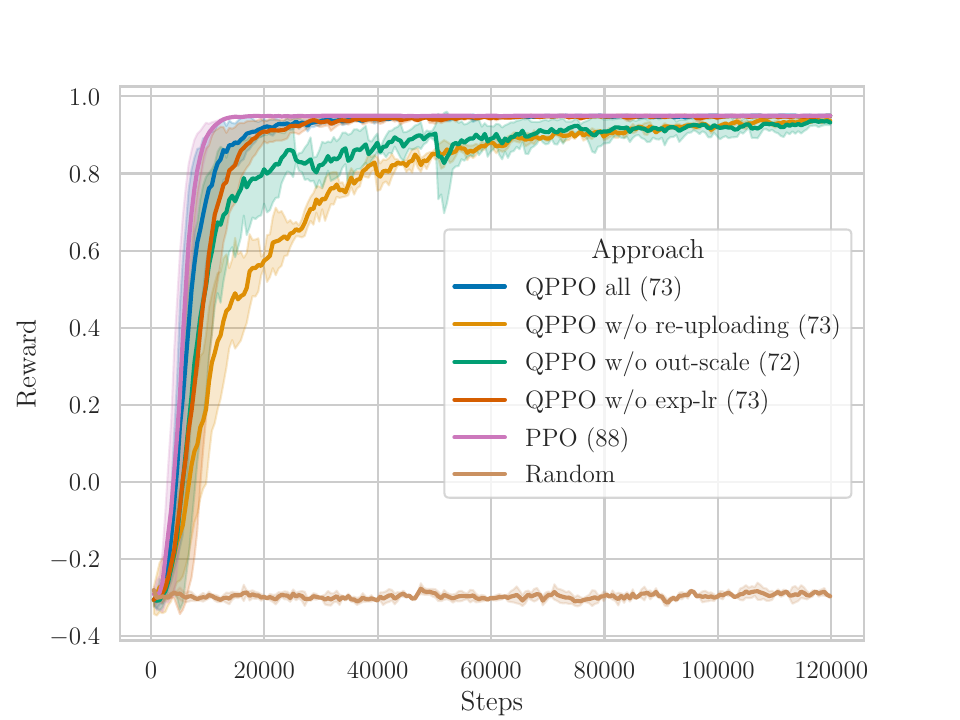}
   \label{fig:FL-Ansatz-Comparison}
  }
  \hspace{0.01\linewidth}
  \subfloat[Local output scaling in Frozen Lake over 25,000 Time Steps.]{
   \includegraphics[width=0.30\linewidth]{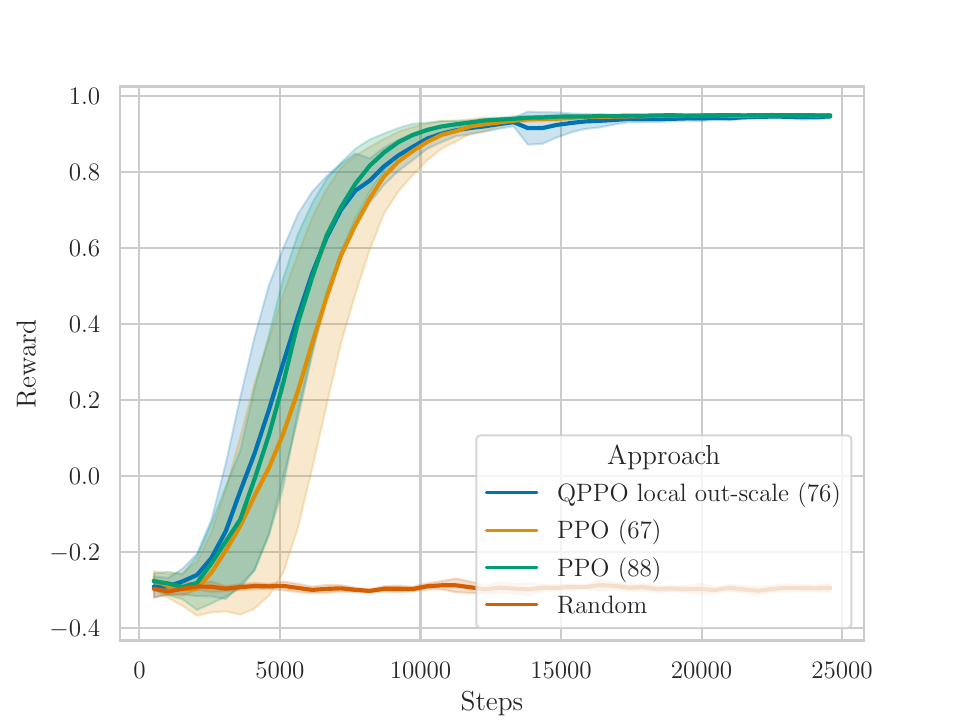}
   \label{fig:combined:b}
  }
  \hspace{0.01\linewidth}
  \subfloat[Standard approach vs. circuits from Kölle \textit{et al.} \cite{Hgog2023} and Jerbi \textit{et al.} \cite{QRL_jerbi2021parametrized} over 50,000 steps.]{
   \includegraphics[width=0.30\linewidth]{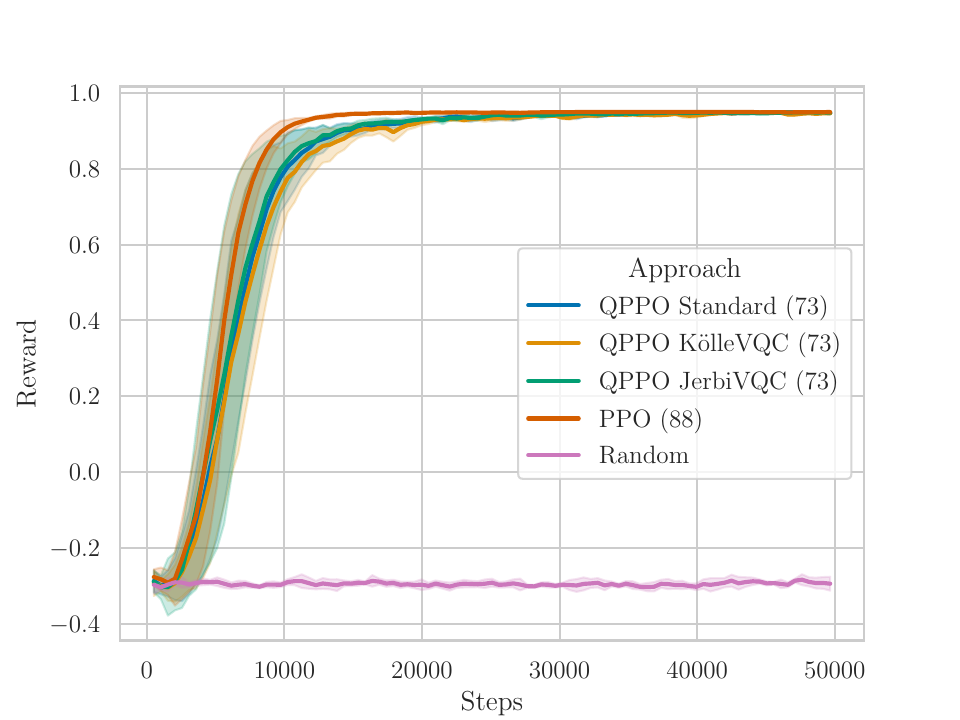}
   \label{fig:combined:a}
  }
  \caption{Comparative analysis across different approaches and environments: (a) evaluates the average reward over three runs using data re-uploading, global output scaling, and exponentially decaying learning rate; (b) illustrates the impact of local output scaling in Frozen Lake allowing for a higher $lr=2.5 \times 10^{-2}$ compared to (a) $lr=5 \times 10^{-3}$; and (c) compares the standard QPPO approach with circuits from Kölle \textit{et al.} \cite{Hgog2023} and Jerbi \textit{et al.} \cite{QRL_jerbi2021parametrized} in Frozen Lake. The number of required actor parameters is shown in parentheses, with all approaches using a critic with 5313 parameters \cite{meyer2022survey}.}
  \label{fig:combined}
\end{figure*}

The initial experiments in the Frozen Lake environment aimed to evaluate the impact of data re-uploading, global output scaling, and the exponentially decaying learning rate introduced in this work on the QPPO's performance. \cref{fig:FL-Ansatz-Comparison} presents the experimental results, displaying the average smoothed outcome over three runs against the number of time steps, with the standard deviation illustrated as transparent areas.

\begin{figure}[htbp]
  \centering
  \includegraphics[width=\linewidth]{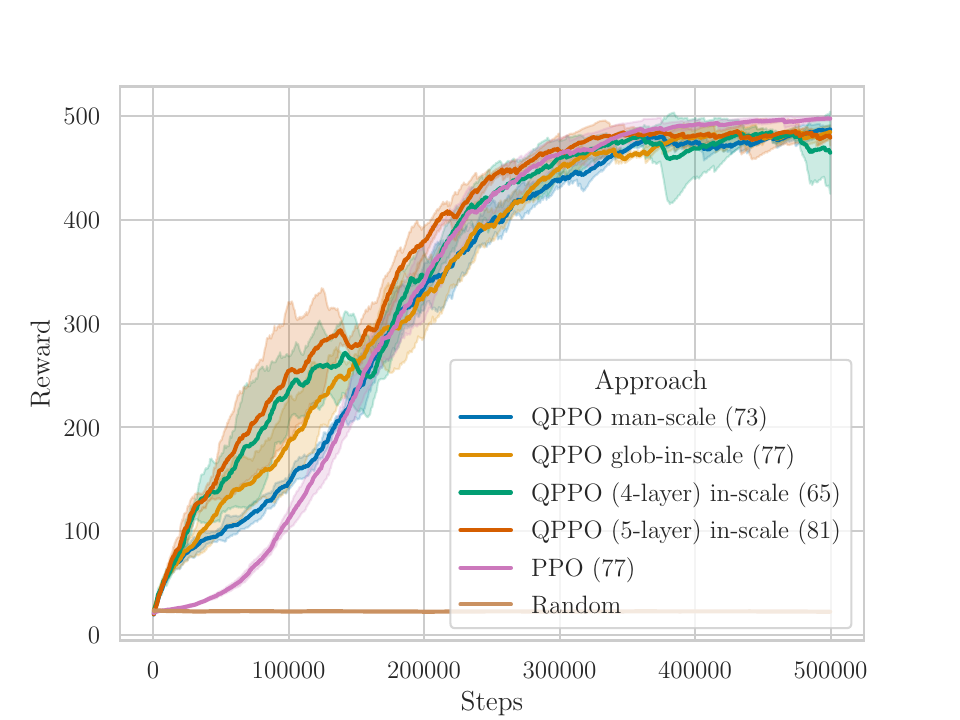}\\
  \caption{
     QPPO with manual re-scaling, global input scaling, input scaling (with 4 and 5 layers), and classical PPO in Cart Pole.}\label{fig:combined:c}
\end{figure}

\begin{figure}[htb]
 \centering
  \subfloat[Frozen Lake]{
   \label{fig:CP_outscale:b}
   \includegraphics[width=0.48\linewidth]{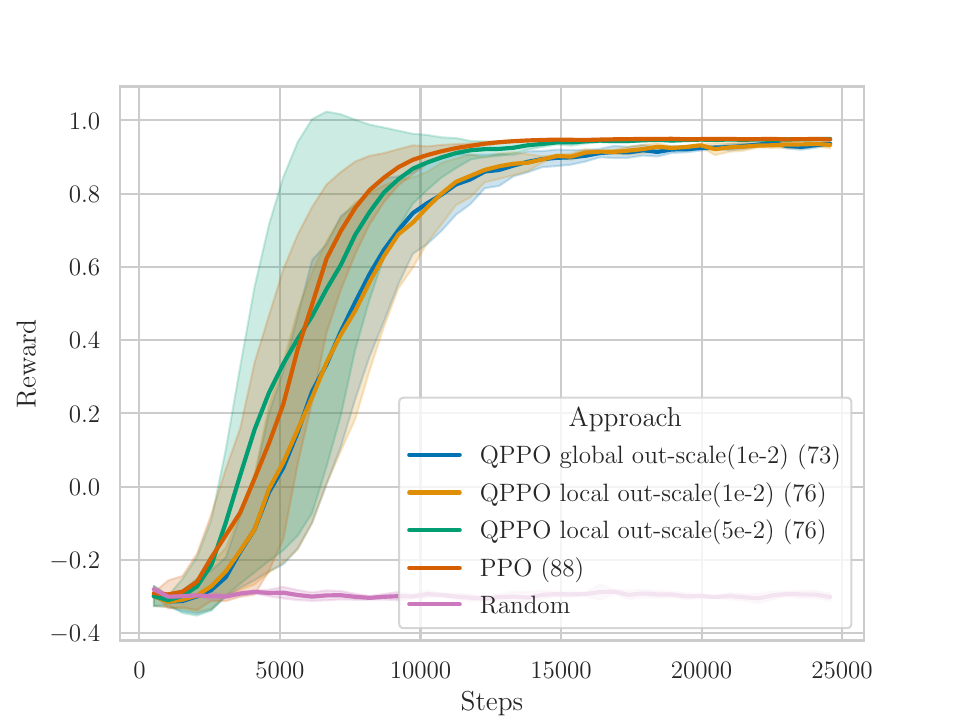}}
    \subfloat[Cart Pole]{
   \label{fig:CP_outscale:a}
   \includegraphics[width=0.48\linewidth]{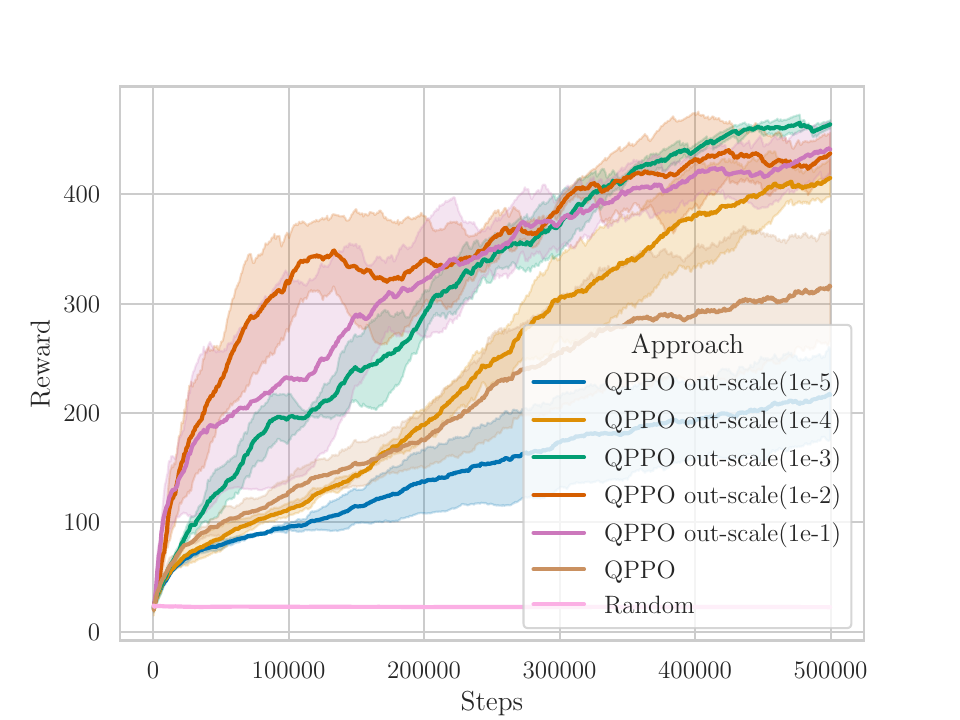}}
 \caption{
  Output scaling strategies comparison. Parentheses contain the scaling learning rate and the number of parameters used for the actor. (a) Frozen Lake: Initially, a ScaleLr of $10^{-2}$ seems optimal, but due to instabilities, $5 \times 10^{-3}$ is later chosen for better reproducibility. (b) Cart Pole: We chose a output scaling learning rate of $10^{-4}$ for further experiment and $2 \times 10^{-4}$ for final tests, to avoid instabilities.
}
 \label{fig:CP_outscale}
\end{figure}

The results in \cref{fig:FL-Ansatz-Comparison} demonstrate that all evaluated techniques positively affect learning success. Each tested approach, successfully learned an optimal strategy for solving the Frozen Lake environment for each seed. However, none matched the performance of classical PPO with the same number of parameters. While PPO converges in approximately 20,000 steps with minimal differences between three and four hidden layers, QPPO with data re-uploading and global output scaling requires about 50,000 steps. As seen in \cref{fig:CP_outscale}), the learning rate plays a critical role when using output scaling.

\begin{figure}[htb]
 \centering
  \subfloat[Learning Rates]{
   \label{fig:CP_qlr:a}
   \includegraphics[width=0.48\linewidth]{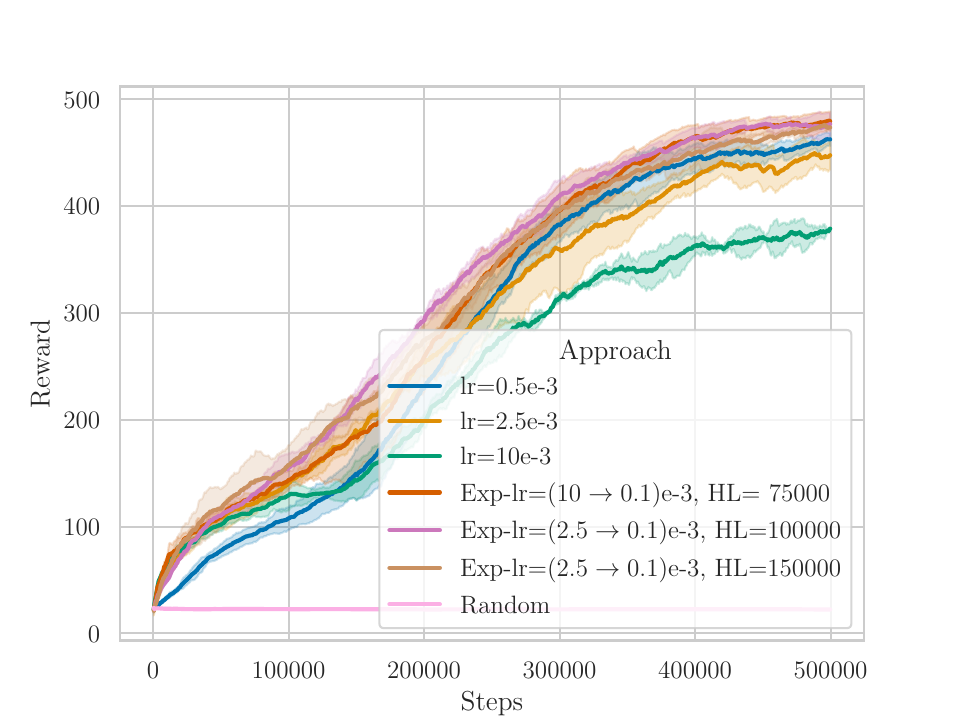}}
  \subfloat[Schedules]{
   \label{fig:CP_qlr:b}
   \includegraphics[width=0.48\linewidth]{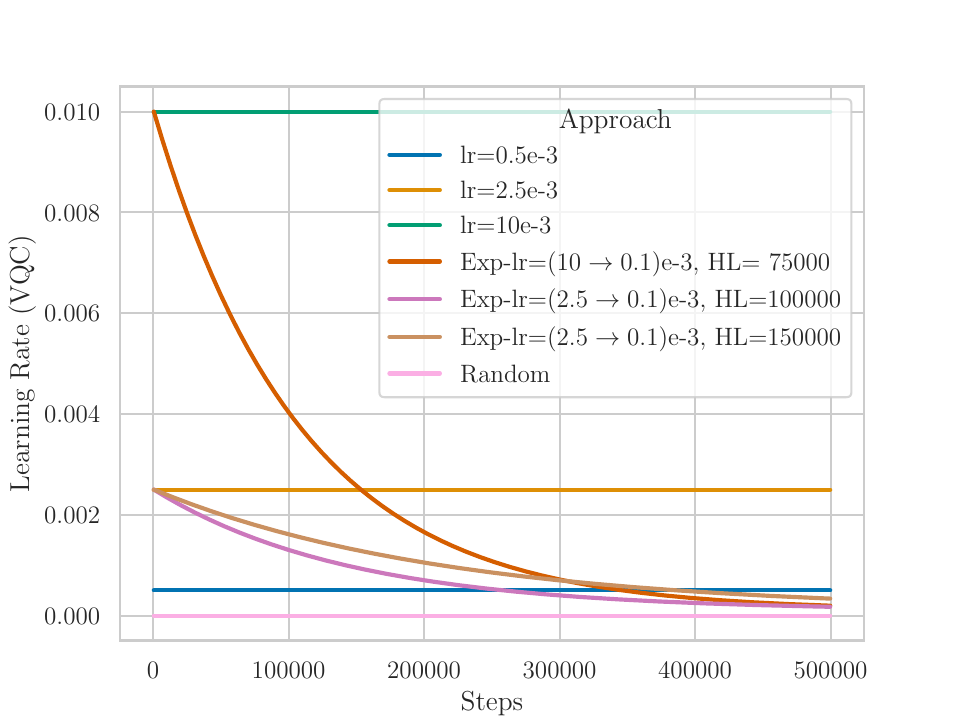}} 

 \caption{Learning rate schedule comparison in Cart Pole. A exponentially decaying learning rate from $2.5$ to $0.1 \times 10^{-3}$ with a half-life of $100000$ is chosen for further tests.}
 \label{fig:CP_qlr}
\end{figure}

Our approach using an exponentially decaying learning rate initially progresses faster than without it, though the difference is less pronounced than during our preliminary experiments in \cref{fig:CP_qlr}. This method consistently offered an advantage for QPPO across all test phases, unlike for classical PPO, where it did not significantly enhance success.

Our experiments (\cref{fig:CP_outscale}) reveal that VQCs lacking output scaling and decaying learning rate exhibit more pronounced convergence issues, tending to converge very slowly at high learning rates or prematurely at low rates to suboptimal strategies. Thus, identifying a consistent VQC learning rate that both optimizes the path and consistently achieves the goal without output scaling proves challenging.

\begin{figure}[htb]
 \centering
  \subfloat[Data re-uploading]{
   \label{fig:CP_data:b}
   \includegraphics[width=0.48\linewidth]{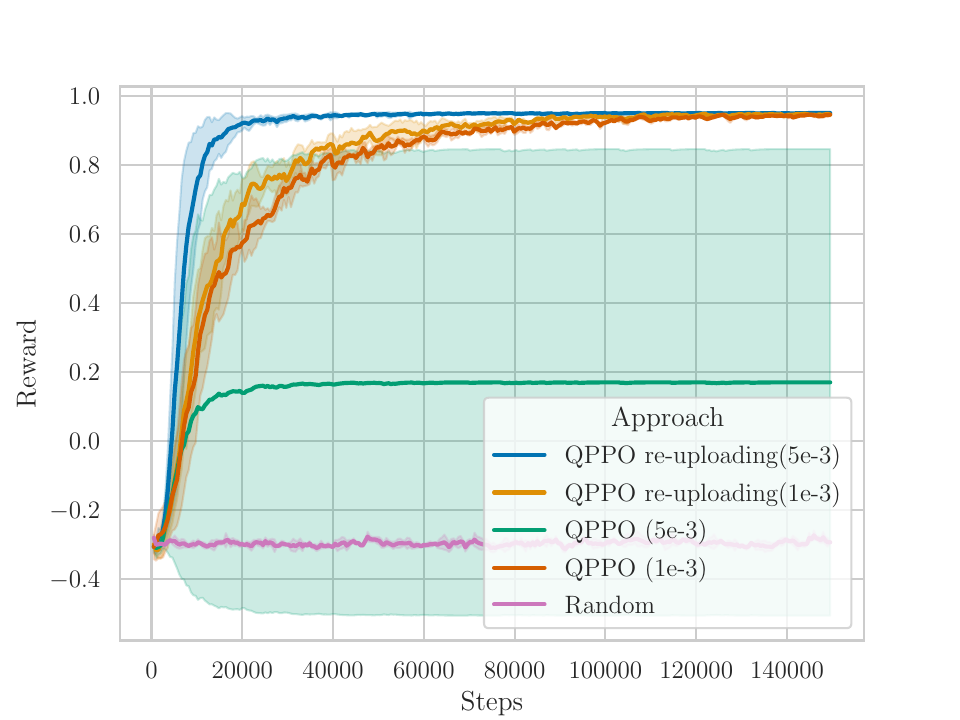}}
    \subfloat[Input scaling]{
   \label{fig:CP_data:a}
   \includegraphics[width=0.48\linewidth]{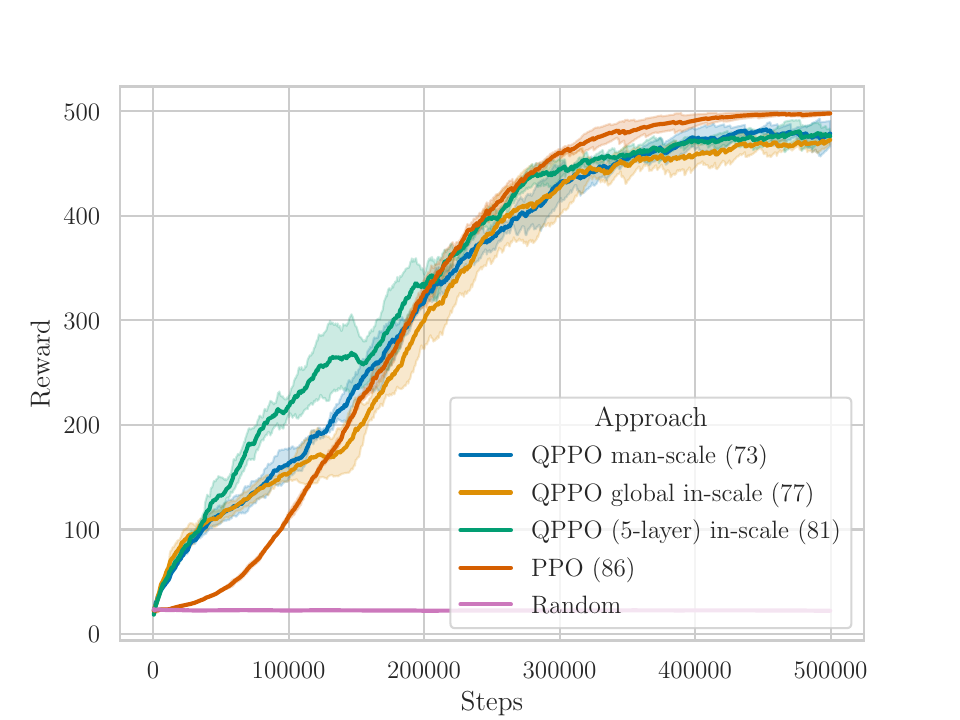}}
 \caption{
 Impact of data re-uploading (Frozen Lake) and input scaling (Cart Pole).
}
 \label{fig:CP_datareup}
\end{figure}

The approach without data re-uploading performed the worst in this test, primarily because it required a significantly lower output scaling learning rate ($10^{-3}$ instead of $5 \times 10^{-3}$) to avoid converging to local minima in some runs. \cref{fig:CP_datareup} confirms a clear disadvantage compared to the approach with re-uploading and the same rate, indicating lower robustness. Despite these challenges, this QPPO approach eventually learns the optimal strategy.

However in \cref{fig:combined:b} and \cref{fig:CP_outscale:b}, we investigated the use local output scaling, where each output is scaled independently. Compared to global output scaling, local output scaling allows for a significantly higher learning rate of $2.5 \times 10^{-2}$ instead of $5 \times 10^{-3}$ before learning becomes unstable. This results in a faster convergence but similar performance.

A follow-up test (\cref{fig:combined:a}) compared the complete standard approach with circuits from the works of Kölle \textit{et al.} \cite{Hgog2023} and Jerbi \textit{et al.} \cite{QRL_jerbi2021parametrized}, adapted for QPPO as described in \cref{subsec:alt_arch}. Despite fundamentally different architectures, the variations in results were negligible, indicating that circuit details are relatively unimportant compared to the performance-enhancing methods evaluated in this work and a good hyperparameter search.


The standard setup, tested again with local output scaling over five runs (\cref{fig:CP_outscale}), permitted a considerably higher scaling learning rate of $2.5 \times 10^{-2}$ without failing any runs, thereby significantly increasing the learning speed to match classical PPO. At lower learning rates, it behaved similarly to global scaling with the same rate.


Finally, we tested manual rescaling, global input scaling (one parameter per input dimension), and (normal or local) input scaling over 500,000 time steps (\cref{fig:combined:c}). To accommodate the additional parameters in the latter approach, we reduced it to 4 or 5 layers (65 or 81 parameters). The results, compared in \cref{fig:CP_data:a}, show all approaches achieve nearly optimal solutions. Shared input scaling performed similarly to manual rescaling, while local input scaling with 4 layers lagged slightly behind. Adding another layer and 16 more parameters slightly exceeded manual rescaling, suggesting comparable performance relative to the number of trainable parameters. Despite these positive results, none of the tested QPPOs surpassed classical PPO, which consistently achieved good results irrespective of parameter count.

\section{Discussion}\label{sec:Discussion}
This study has demonstrated the significance of recent literature modifications on the performance of a VQC utilized in a QPPO algorithm. Consistent with findings from \cite{perez2020data_re-uploading} and \cite{QRL_jerbi2021parametrized}, we confirmed that data re-uploading facilitates faster and more stable learning by enhancing the VQC's sensitivity to environmental states. While an unmodified VQC struggled to consistently find and reach the shortest path in the simple Frozen Lake environment, introducing output scaling effectively resolved this issue through efficient greediness control. Although a wide range of learning rates for the scaling factor is viable for basic circuit learning capability, optimizing this hyperparameter is crucial for the learning speed of stochastic policies, as excessively high values can derail the learning process, and too low values can severely delay progress.

Local scaling achieved slightly superior results in Frozen Lake with a higher learning rate, though this appears to be a specific case. Since global and local scaling perform equivalently in most settings, scaling with a parameter per dimension achieves greater robustness by biasing towards unnecessary actions. In the deterministic environment of Frozen Lake, higher greediness in action selection generally improves the learning curve, a condition not commonly present in other, particularly stochastic, environments where actions are needed almost equally. Thus, global scaling, with fewer parameters, is generally preferable.

The newly introduced exponentially decaying learning rate for VQCs further improved circuit success and stability, especially for the unmodified VQC. Conversely, input scaling was only effective in reducing the number of layers while maintaining potential and equivalent parameter count, showing no benefits on performance per parameter. Overall, the modified QPPO achieved performance close to classical PPO with the same number of parameters but did not surpass it despite improvements.

The hypothesis that VQCs, due to their expressive power relative to NNs with the same number of parameters, would perform better, and the tested modifications would contribute valuable enhancements. This was confirmed for all modifications except input scaling. Specifically, data re-uploading, as corroborated by \cite{perez2020data_re-uploading}, enhanced VQC performance without adding more parameters. Similarly, output scaling significantly increased VQC learning capability and largely eliminated exploration-exploitation trade-off issues with just one additional parameter. This finding is novel as \cite{QRL_skolik2022quantum} demonstrated the relevance of output scaling only in value-based algorithms, and \cite{QRL_jerbi2021parametrized}, despite acknowledging the importance of softmax output for achieving good results, barely discussed the trainable parameter's influence or its critical impact on stochastic policy success. Further success in QPPO was achieved by adopting an appropriate, exponentially decaying learning rate over the learning process, a strategy already common in some PPO implementations but newly discovered to behave significantly differently for VQCs compared to NNs, suggesting a rapidly decaying exponential learning rate as more suitable for VQCs.

The fact that input scaling did not reduce parameter count, while not surprising given \cite{QRL_jerbi2021parametrized}'s aim for a genuine quantum advantage in a specially designed environment without focusing on parameter count, was nonetheless disappointing. However, it at least demonstrated that effectiveness does not deteriorate with fewer layers. Since approaches utilizing input scaling initially learned much faster than those without it (without any hyperparameter optimization specifically for this approach), dismissing its advantage outright may be premature.

Finally, the expectation that these methods would suffice to outperform classical PPO was not confirmed. This contrasts with studies like \cite{QRL_acuto2022variational} and \cite{QRL_lan2021variational}, which showed Quantum Soft Actor Critic algorithms matching or exceeding classical versions, suggesting a similar possibility for QPPO. This discrepancy might be due to a lack of exhaustive hyperparameter search, relying instead on standard PPO settings not tailored for VQC use, possibly giving NNs a slight advantage. Alternatively, other algorithms like Soft Actor Critic might benefit more from VQC's expressive power than PPO.

Covering both discrete (Frozen Lake) and continuous (Cart Pole) state and action environments, and ensuring reproducible results through tests across at least three runs, this study nonetheless acknowledges the potential influence of initialization randomness and the absence of exhaustive hyperparameter search due to the lengthy quantum simulations. However, the clear differences between QPPO versions and the negligible impact of VQC architecture on outcomes suggest that the positive effects of the tested methods could likely be reproduced with other (VQC-based) quantum algorithms.

\section{Conclusion}\label{sec:Conclusion}

In this work, we aimed to provide deeper insights into some of the most recent methods, evaluating that data re-uploading, output scaling, and the newly introduced exponentially decaying learning rate for VQCs all contribute towards the goal of parameter reduction. However, our experiments indicated that these modifications alone are insufficient to surpass the efficiency of classical PPO. The rational conclusion from our findings recommends incorporating data re-uploading into all future QRL endeavors, as it offers a significant benefit without adding more parameters and is trivial to implement. It was also demonstrated that output scaling is fundamental for policy-based methods, suggesting that VQCs in more complex environments are likely going to fail without some form of greediness control. However, anyone wishing to utilize trainable output scaling to regulate greediness must be aware of its strong influence on the stochastic policy and its co-dependence on the general learning rate of the parameterized circuit, for which a satisfactory optimization approach has yet to be found. Additionally, it was shown that an exponentially decaying learning rate positively impacts the VQC, enhancing hyperparameter stability. Lastly, while input scaling was confirmed as an efficient method to increase the expressiveness of the VQC with the same number of layers, it was acknowledged that it does not contribute to the fundamental goal of parameter reduction. These insights should provide a stable knowledge base for future researchers to decide whether these methods are suitable for their VQC, thus bringing closer the goal of developing an even more efficient algorithm with a provable quantum advantage.

In the future, testing the QPPO for continuous actions would be intriguing to determine if the VQC can excel in more complex state-action spaces. This is particularly relevant since the greediness regulation of input scaling, found to be fundamental for the approach, does not directly translate to continuous actions, necessitating an intelligent solution to avoid similar convergence issues. Additionally, executing the approach on a real quantum computer to see if it can still compete with the classical approach despite quantum noise would be valuable. If the focus shifts less towards gaining new insights and more towards further optimizing the QPPO to surpass the classical version, incorporating amplitude encoding could significantly reduce the required number of qubits (and thus parameters). Once the hyperparameters for the actor are fully optimized, replacing the critic with a VQC would be a logical next step, which was not explored in this work due to time constraints.

\section*{Acknowledgements}
This work is part of the Munich Quantum Valley, which is supported by the Bavarian state government with funds from the Hightech Agenda Bayern Plus.

\section*{Appendix}

\begin{figure}[htb]
 \centering
  \subfloat[Frozen Lake]{
   \label{fig:CP_inits:a}
   \includegraphics[width=0.48\linewidth]{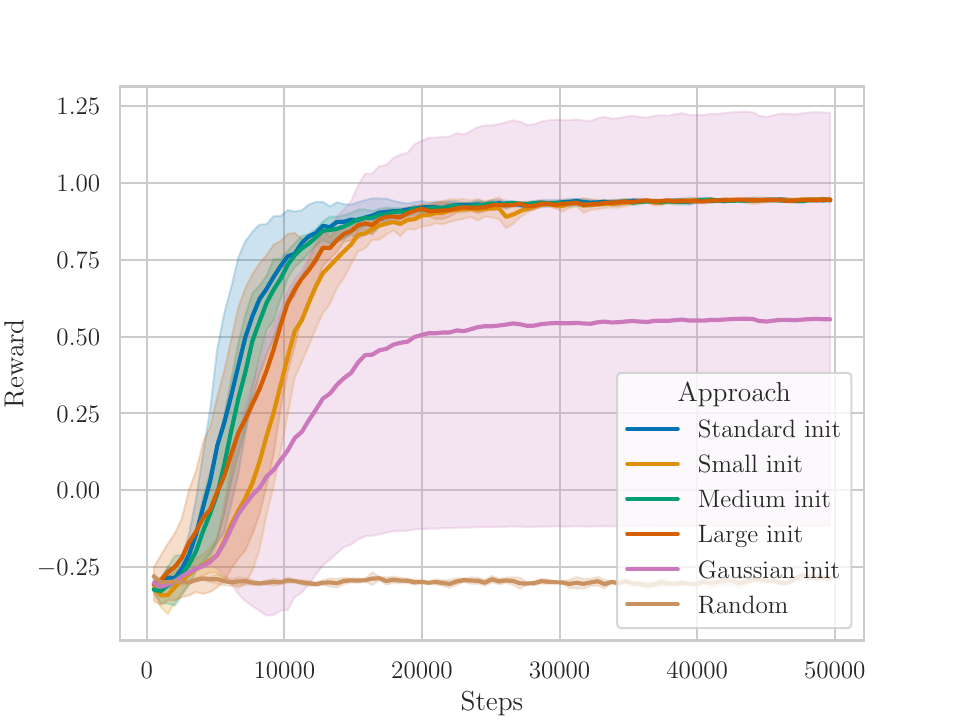}}
  \subfloat[Cart Pole]{
   \label{fig:CP_inits:c} 
   \includegraphics[width=0.48\linewidth]{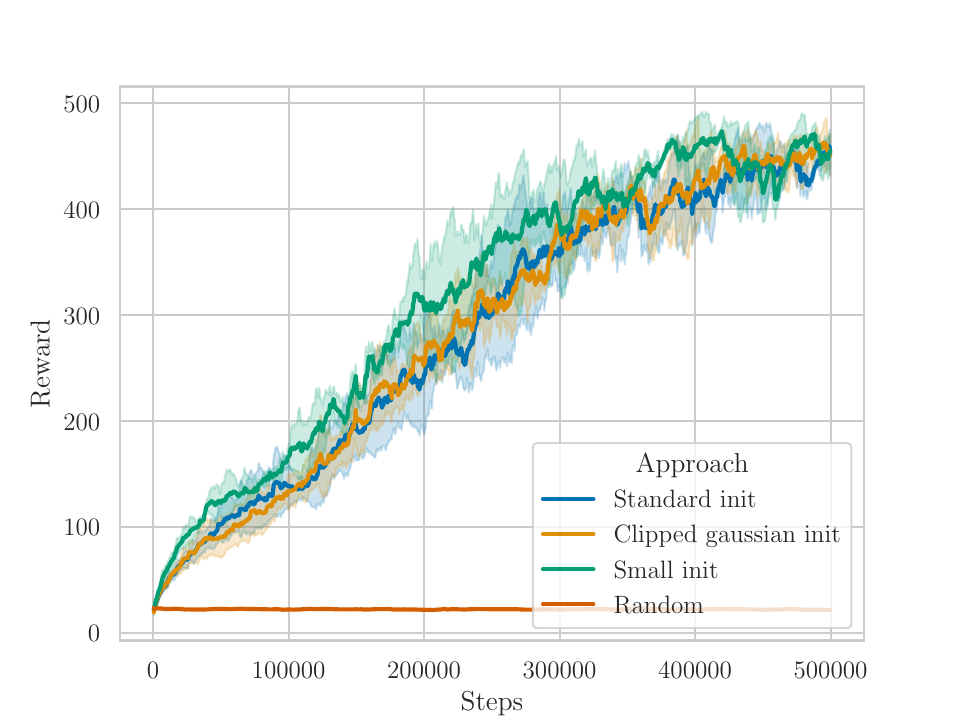}}

 \caption{Parameter initialization strategies comparison. (a) Frozen Lake: Standard initialization performs best and is used for further tests. (b) Cart Pole: Small initialization achieves the best result and is used for all further tests.}
 \label{fig:CP_inits}
\end{figure}

\begin{table}[htb]
  \centering
  \begin{tabularx}{\linewidth}{|X|l|}
    \hline
    Parameter Name                & Value                 \\ \hline\hline
    Batchsize                     & 512                  \\ \hline
    Number of Minibatches         & 4                    \\ \hline
    Mini-Batch-Size               & 128                  \\ \hline
    Cycles per update             & 4                    \\ \hline
    GAE-$\lambda$                 & 0.95                 \\ \hline
    Discount Factor $\gamma$      & 0.99                 \\ \hline
    Clip Coefficient              & 0.2                  \\ \hline
    Value Loss Coefficient        & 0.5                  \\ \hline
    Entropy Loss Coefficient      & 0.01                 \\ \hline
    Max Gradient Norm             & 0.5                  \\ \hline
    Critic NN Hidden Layers       & 2                    \\ \hline
    Critic NN Hidden Layer Nodes  & 64                   \\ \hline
    Critic NN lr                  & $2.5 \times 10^{-4}$ \\ \hline
    VQC Variational Layers        & 6                    \\ \hline
    VQC Encoding Layers           & 6                    \\ \hline
    Adam's-$\epsilon$             & $10^{-5}$            \\ \hline
  \end{tabularx}
  \caption{PPO Hyperparameters}
  \label{tab:general_hyperparams}
\end{table}

\begin{table}[htb]
  \centering
  \begin{tabularx}{\linewidth}{|l|X|X|}
    \hline
    Parameter Name                  & Frozen Lake                                               & Cart Pole                 \\ \hline\hline
    VQC Fixed LR                    & $2.5 \times 10^{-3}$                                      & $5 \times 10^{-4}$       \\ \hline
    VQC Exp-LR Start                & $10^{-2}$                                                 & $2.5 \times 10^{-3}$     \\ \hline
    VQC Exp-LR End                  & $10^{-4}$                                                 & $10^{-4}$                \\ \hline
    Half-life                       & $25,000$                                                  & $100,000$                \\ \hline
    Global Output Scaling LR        & $5 \times 10^{-3}$ ($10^{-3}$ w/o re-uploading)  & $2 \times 10^{-4}$       \\ \hline
    Local Output Scaling LR         & $2.5 \times 10^{-2}$                                      & N/A                      \\ \hline
    Parameter Initialization        & Standard                                                  & Small                    \\ \hline
    Actor NN LR                     & $10^{-2}$                                                 & $10^{-4}$                \\ \hline
    Time Steps                      & $150,000$                                                 & $500,000$                \\ \hline
    EWMA-$\alpha$                    & $0.3$                                                     & $0.015$                  \\ \hline
  \end{tabularx}
  \caption[Optimized Hyperparameters for Frozen Lake and Cart Pole]{
    Optimized Hyperparameters for Frozen Lake and Cart Pole
  }\label{tab:merged_hyperparams}
\end{table}

\clearpage

\bibliographystyle{unsrt}  
\bibliography{main} 
\end{document}